\documentclass[twocolumn]{aastex63}

\usepackage{graphicx}	
\usepackage{amsmath}	
\usepackage{amssymb}	
\usepackage{latexsym}
\usepackage{hyperref}	
\usepackage{mathtools}
\usepackage{rotating}
\usepackage{epsfig}

\newcommand{\R}{\mathbb{R}}                  
\newcommand{\N}{\mathbb{N}}
\newcommand{\Z}{\mathbb{Z}}
\newcommand{\C}{\mathbb{C}}

\received{--}
\revised{--}
\accepted{--}
\submitjournal{AJ}

\shorttitle{The dynamics of asteroid Kamo'oalewa}
\shortauthors{Fenucci and Novakovi\'c}

\begin{document}

\title{The role of the Yarkovsky effect in the long-term dynamics of asteroid (469219) Kamo'oalewa}

\correspondingauthor{Marco Fenucci}
\email{marco\_fenucci@matf.bg.ac.rs}

\author[0000-0002-7058-0413]{Marco Fenucci}
\affiliation{Department of Astronomy, Faculty of Mathematics \\
University of Belgrade \\ 
Studentski trg 16, 11000 Belgrade, Serbia}

\author[0000-0001-6349-6881]{Bojan Novakovi\'c}
\affiliation{Department of Astronomy, Faculty of Mathematics \\
University of Belgrade \\ 
Studentski trg 16, 11000 Belgrade, Serbia}










\begin{abstract}
Near-Earth asteroid (469219) Kamo'oalewa (aka 2016 HO3) is an Earth co-orbital and a potential space mission target. Its short-term dynamics is characterized by a periodic switching between quasi-satellite and horseshoe configurations. Due to its small diameter of only about 36 meters, the Yarkovsky effect may play a significant role in the long-term dynamics. In this work, we addressed this issue by studying the changes in the long-term motion of Kamo'oalewa caused by the Yarkovsky effect. We used an estimation of the magnitude of the Yarkovsky effect assuming different surface compositions and introduced the semi-major axis drift by propagating orbits of test particles representing the clones of the nominal orbit. Our simulations showed that the Yarkovsky effect may cause Kamo'oalewa to exit from the Earth co-orbital region a bit faster when compared to a purely gravitational model. Nevertheless, it still could remain an Earth companion for at least 0.5 My in the future. Our results imply that Kamo'oalewa is the most stable Earth's co-orbital object known so far, not only from a short-term perspective but also on long time scales.
\end{abstract}



\section{Introduction}

Earth's \emph{quasi-satellites} are objects in a specific type of co-orbital configuration with our planet, such that they stay close to the Earth over a period of time. In contrast to true satellites, orbits of quasi-satellites are located outside the Earth gravitational sphere of influence and therefore are typically dynamically unstable.

\begin{figure}
   \centering
   \includegraphics[width=0.45\textwidth]{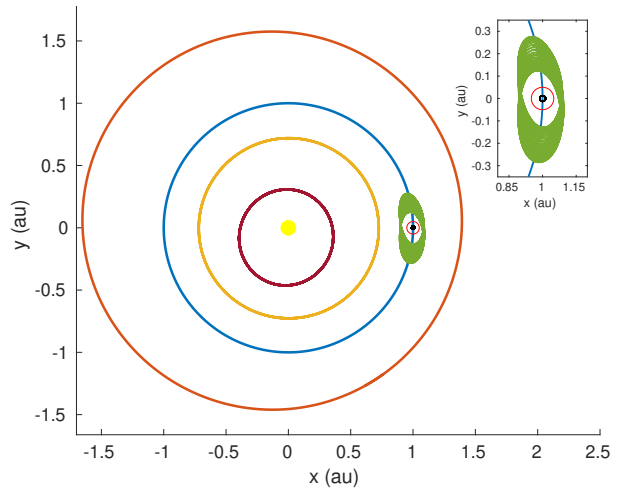}
   \caption{The motion of Kamo'oalewa (green curve) in the time interval $(0
      \text{ yr}, \, 100\text{ yr})$, projected on the ecliptic plane in a heliocentric
      frame that rotates with the Earth. The central yellow dot indicates the Sun, while
   the brown, yellow, blue, and red curves are the orbits of Mercury, Venus, the Earth, and Mars, respectively. 
   The smaller frame shows a magnification of the dynamics around the Earth, where the orbit of the Moon in black and the Hill radius of the Earth in red are plotted, both expanded by a factor 5.}
   \label{fig:orbit}
\end{figure}

(469219) Kamo'oalewa is a small Earth quasi-satellite, that has been discovered by the Pan-STARRS 1 survey telescope at Haleakala Observatory on  2016 April 27. Its name is a composition of Hawaiian words (\textit{ka} (the), \textit{mo'o} (fragment),
\textit{a} (of) and \textit{lewa} (to oscillate)\footnote{Source: NASA/JPL \url{https://ssd.jpl.nasa.gov/sbdb.cgi?sstr=3752445\#content}}), and it has been chosen according to its particular orbital motion, which appears to oscillate for an observer placed on the Earth, as illustrated in Fig.~\ref{fig:orbit}.

Due to its proximity to the Earth and its orbital stability, Kamo'oalewa is a
good candidate for future in-situ exploration and test of new technologies for
asteroid redirection and mining \citep{niklas-emami_2018}. For these reasons, it has attracted lots of attention from scientists, and two missions have already been
proposed so far: the first one, the Near-Earth Asteroid Characterization and Observation
\citep[NEACO;][]{venigalla-etal_2019}, is a concept study aimed to map the surface, develop a shape model,
estimate the mass, and understand the composition of the surface of Kamo'oalewa; the second one,
the ZhangHe, has been designed by the Chinese Academy of Space Technology and it
consists in a sample-return mission to Kamo'oalewa and the subsequent exploration of the main-belt
comet 311P/PANSTARRS\footnote{The original plan was to visit main-belt comet 133P/Elst-Pizarro, but the second target has been changed recently.}. To this purpose, simulations of the performances of the radio-science experiment aiming to estimate the gravitational field for the ZhangHe mission have been recently carried out by \cite{jin-etal_2020}.

The most complete analysis of the dynamical stability of Kamo'oalewa so far has been performed by \citet{delafuente-delafuente_2016}. The authors found that it is dynamically the most stable Earth's quasi-satellite known so far and that it could remain a companion of our planet for more than 1 Myr.
In this work, however, \citet{delafuente-delafuente_2016} did not consider the possible role of non-gravitational effects.

Non-gravitational forces, and in particular the Yarkovsky effect \citep[see e.g.][]{bottke-etal_2006}, may have important consequences for the long-term dynamics of Kamo'oalewa, and reliable conclusions about the stability could be drawn only taking into account both gravitational and non-gravitational effects. This may be especially true for such small objects as Kamo'oalewa (see Sec.~\ref{sec:basic}). In this respect, though theoretical predictions cast some doubts on the feasibility of the large magnitude of the Yarkovsky effect for small asteroids, recent observational detection showed that resulting semi-major axis drift induced by the Yarkovsky effect could be very fast \citep{greenberg-etal_2020}.

In this paper, we extended previous works on the dynamical stability of Kamo'oalewa by addressing the role of the Yarkovsky effect.
Accurate modeling of the Yarkovsky effect requires knowledge of the physical properties of the object \citep{vokrouhlicky-etal_2015}, which are mostly still not available. However, the semi-major axis drift induced by the Yarkovsky effect may be successfully estimated using statistical methods, like the one recently developed by \citet{fenucci-etal_2021}, that we adopted here.

\section{A short review of recent literature}
\label{sec:basic}

As a near-Earth object (NEO), Kamo'oalewa 
belongs to the Apollo group.
The distance from our planet stays between 38 and 100 lunar distances 
\citep{delafuente-delafuente_2016}, well beyond the Hill sphere of influence of the Earth, that
is about 3.9 lunar distances wide. Moreover, the heliocentric orbit is currently in 1:1 mean motion 
resonance with the Earth, with critical angle $\sigma = \lambda-\lambda_E$ librating around zero, 
where $\lambda, \lambda_E$ are the mean longitudes\footnote{By mean longitude we refer to the angle $\lambda = \Omega + \omega + \ell$.}.
Due to the particular behaviour of $\sigma$, Kamo'oalewa is also classified as a
quasi-satellite. So far, only five Earth's quasi-satellites are known \citep{jedicke-etal_2018},
but Kamo'oalewa is the closest and the most stable one.

\citet{delafuente-delafuente_2016} studied both the short and the long-term dynamics of Kamo'oalewa, finding that it switches repeatedly between the quasi-satellite configuration
(i.e. critical angle $\sigma$ librating around 0$^\circ$), and the horseshoe configuration
(i.e. critical angle $\sigma$ librating around 180$^{\circ}$, with an amplitude of almost 180$^{\circ}$), experiencing about 24 quasi-satellite events during 100 ky of numerical
simulations. The current quasi-satellite configuration started about 100 years ago, and it
will last for at least other 300 years. 
During the quasi-satellite episodes, the value of the eccentricity $e$ is the highest, while
the one of the inclination $i$ the lowest. Moreover, the switching takes place when the 
descending node is farthest
from the Sun and the ascending node is closest. The value of the Lidov-Kozai parameter
$\sqrt{1-e^2}\cos i$ remains fairly constant, and the eccentricity, the inclination and
the longitude of the ascending node oscillate with the same frequency. However, a better
study of the coupled oscillations revealed that Kamo'oalewa is not in a Lidov-Kozai
resonance. 

Some physical characteristics have been studied as well, and preliminary values have been determined.
\citet{reddy-etal_2017} used observations from the
Large Binocular Telescope (LBT) and the Discovery Channel Telescope (DCT) to produce a lightcurve, that suggests a rotation period of about 0.467 hr,
corresponding to about 28 minutes. In addition, a low-resolution spectrum in the 0.39-0.97 $\mu$m wavelength band was obtained. The spectrum shows a sharp rise between 0.4 and 0.7 $\mu$m, with a flat profile at higher wavelengths. This is not enough to uniquely determine the taxonomic 
class of Kamo'oalewa, since the comparison with different spectra suggests
it could be either S-type or L-type, with the latter being more likely.
Finally, a diameter of 36 meters has been derived, using an absolute magnitude of $H=24.3$ mag and a typical albedo for S-type asteroids of 0.25. Since the derived rotation period and the spectrum are not uncommon among small NEOs, it has been confirmed that Kamo'oalewa is a natural object rather than space junk.

Recently, \citet{li-scheeres_2020} developed a possible shape model, based on the lightcurve
measurements by \citet{reddy-etal_2017}. 
Using an ellipsoidal model, the authors suggested that Kamo'oalewa is an elongated object, with a ratio between the
shortest and longest axes less than 0.4786. Assuming a value of 2700 kg m$^{-3}$ for
the density, the authors also found that the sphere of influence relative to the Sun is about 150 meters 
on average, containing therefore the entire object. Additionally, studying the cohesive forces,  \citet{li-scheeres_2020} found
that it should be possible for Kamo'oalewa to retain millimeter to centimeter-sized grains on its
surface, preferably in polar and short-axis regions.

\section{Methods}
\subsection{Numerical model and initial conditions}
\label{ss:numericalModel}
Numerical integrations are performed with the $N$-body code
\texttt{mercury}\footnote{\url{https://www.astro.keele.ac.uk/~dra/mercury/}} by J.
Chambers \citep{chambers-migliorini_1997, chambers_2012}, using the Bulirsch-Stoer
integration algorithm \citep{bulirsch-stoer_2002}.
The dynamical model used includes the gravitational attraction of the Sun, the perturbations of the eight
planets from Mercury to Neptune, and the Moon. The code has been modified in order to include also the Yarkovsky 
effect in the equations of motion (see Sec.~\ref{ss:yarkovskyDrift}).  For the purpose of this investigation, the Yarkovsky effect 
is modeled as a secular force along the orbital velocity of the asteroid. The drift in au My$^{-1}$ is given
in input, and it is maintained constant over the whole integration timespan. 
Limitations of this approach have been discussed in Sec.~\ref{sec:discussion}.
\begin{table}
   \centering
   \caption{The nominal osculating orbital elements of Kamo'oalewa and their corresponding uncertainties at epoch 2459000.5 JD, taken from
      NEODyS.}
   \label{tab:orb_el}
   \begin{tabular}{cccc}
      \hline
      \hline
      Orbital     & Value  & $1\sigma$ uncertainty   & Units\\
      element     &   &   & \\
      \hline
      $a$   &  \hphantom{00}1.00124795599 & $1.610 \times 10^{-9}$    &	au\\
   $e$      &  \hphantom{00}0.10330139523 & $1.169 \times 10^{-7}$   & / \\
   $i$      &  \hphantom{00}7.78532901460 & $8.305 \times 10^{-6}$    & deg\\
   $\Omega$ & \hphantom{0}66.15568136854  & $1.074 \times 10^{-5}$    & deg\\
   $\omega$ & 306.18786126354             & $1.825 \times 10^{-5}$    & deg\\
   $\ell$   & 236.29311427075             & $2.433 \times 10^{-5}$    & deg\\
      \hline
   \end{tabular}
\end{table}

Initial conditions for all the massive bodies at epoch 2459000.5 JD (corresponding to 2020 May 31) 
are taken from the JPL Horizons\footnote{\url{https://ssd.jpl.nasa.gov/?horizons}} ephemeris system, and they are
based on the DE405 planetary orbital ephemeris. The nominal orbit, reported
in Table~\ref{tab:orb_el}, and the corresponding covariance matrix for the asteroid Kamo'oalewa 
are taken from the NEODyS\footnote{\url{https://newton.spacedys.com/neodys/index.php?pc=0}} service.

\subsection{Orbital clones}
\label{ss:orbitalClones}
Small uncertainties in the initial orbit can cause large differences in the long-term
numerical integrations, because of the chaotic nature of the $N$-body problem.
To overcome this, it is necessary to take into account the effect of small perturbations
in the initial condition, integrating orbital clones that are compatible with the nominal
orbit obtained from astrometry.

We use the following method to produce the clones. Let us denote with
$\mathbf{x}^* \in \R^6$ the nominal orbit and with $\Gamma \in \R^{6 \times 6}$
the corresponding covariance matrix. The probability density of the
nominal orbit is a six dimensional Gaussian distribution, with average at $\mathbf{x}^*$ and
covariance equals to $\Gamma$ \citep[see e.g.][]{milani-gronchi_2009}, hence we have to generate clones
accordingly. To this end, note that if $\mathbf{u}^* \in \R^6$ is distributed
according to a multidimensional Gaussian with zero average and identity covariance matrix,
then the linear transformation
\begin{equation}
   A\mathbf{u}^* + \mathbf{x}^*
   \label{eq:transformationGaussian}
\end{equation}
defines a Gaussian distribution with average $\mathbf{x}^*$ and covariance matrix $\Gamma$.
The matrix $A \in \R^{6 \times 6}$ is obtained from the Cholesky decomposition
\citep[see e.g.][]{bulirsch-stoer_2002} of $\Gamma$, i.e. 
\begin{equation}
   \Gamma = AA^T,
   \label{eq:choleskyDec}
\end{equation}
where $A$ is lower triangular.
The random vectors $\mathbf{u}^*$ with standard normal distribution are generated using the Box-Muller
algorithm \citep{box-muller_1958}, then clones are created according to Eq.~\eqref{eq:transformationGaussian}.

\subsection{Modeling the Yarkovsky drift}
\label{ss:yarkovskyDrift}
The Yarkovsky effect is a thermal force caused by sunlight \citep{rubincam_1995,
rubincam_1998, farinella-etal_1998, bottke-etal_2006, vokrouhlicky-etal_2015} that affects the
motion of asteroids smaller than about 30 kilometers in diameter. 
Objects are heated up by the Sun, and then re-radiate the energy away in the thermal waveband, causing a tiny but continuous thrust. The re-radiation is delayed due to the non-zero thermal inertia of asteroids and consequently, as objects rotate, the direction of the thrust is not parallel to the heliocentric position vector, but slightly rotated towards the orbital velocity direction. 
As a consequence, the orbit of a prograde rotator enlarges, while it shrinks for retrograde rotators.

Including this thermal effect in the dynamics and estimating its magnitude are fundamental parts 
of the study of the long-term evolution of small bodies.

A simple analytical model of the Yarkovsky effect can be obtained assuming some
simplifications: i) a spherical shape for the object, ii) a circular orbit around the Sun, iii) a
linearization of the surface boundary condition, and iv) a fixed rotation axis. In this
context, the Yarkovsky effect causes only drift in the semi-major axis, and it
has two distinct contributions, namely the diurnal component $(da/dt)_{\text{d}}$ and the seasonal component
$(da/dt)_{\text{s}}$. The total drift $da/dt$ is then obtained summing up the two
different contributions. Moreover, the total magnitude depends on several orbital and
physical characteristics of the asteroids: the semi-major axis $a$, the diameter $D$, the
density $\rho$, the thermal conductivity $K$ and the heat capacity $C$ of the surface, the
obliquity $\gamma$, the rotation period $P$, the absorption coefficient $\alpha$ and the
surface emissivity $\varepsilon$. The analytical expressions of $(da/dt)_{\text{d}}$ and
$(da/dt)_{\text{s}}$ can be found, for instance, in \citet{vokrouhlicky_1998, vokrouhlicky_1999}.

To understand the typical drift rates that Kamo'oalewa can achieve, we evaluate
\begin{equation}
   \left( \frac{da}{dt} \right)(a,D,\rho,K,C,\gamma,P,\alpha,\varepsilon),
   \label{eq:yarkoForce}
\end{equation}
where the parameters have to be chosen according to the known properties of the object.
However, physical characteristics are usually known up to a certain degree, and some 
of them may be even unknown. For this reason, following \citet{fenucci-etal_2021}, we modeled the input parameters and used a Monte Carlo approach
to predict the most likely values of the semi-major axis drifts given by
Eq.~\eqref{eq:yarkoForce}.

\subsubsection{Modeling the input parameters}
\label{ss:parameterModeling}
We follow a probabilistic model of the parameters similar to \citet{fenucci-etal_2021}, that we briefly describe hereafter.

\textbf{Semi-major axis.} 
The semi-major axis determined from astrometry has a very small uncertainty, of the order of $10^{-9}$ au (see Table~\ref{tab:orb_el}), which does not introduce any relevant error in the computation of the estimated Yarkovsky drift of Eq.~\eqref{eq:yarkoForce}. Hence, we use the nominal value for this parameter. 

\begin{figure*}
   \centering
   \includegraphics[width=\textwidth]{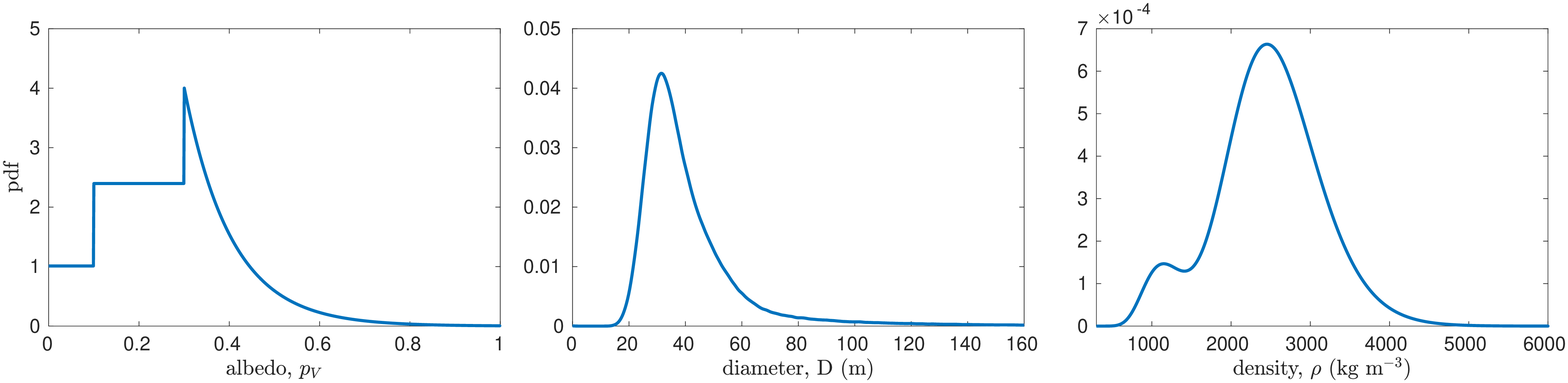}
   \caption{Probability density function for the albedo $p_V$ (left panel), the diameter $D$ (central panel), and the density $\rho$ (right panel), computed for the asteroid Kamo'oalewa.}
   \label{fig:pdf_input}
\end{figure*}

\textbf{Diameter and density.} A population-based distribution for 
the diameter $D$ and the density $\rho$ of a given NEO has been developed in
\citet{fenucci-etal_2021}, and it is based on the NEOs orbital and absolute magnitude distribution model
by \citet{granvik-etal_2018} and the NEOs albedo distribution by \citet{morbidelli-etal_2020}. 
The NEOs orbital distribution model provides the probability for Kamo'oalewa to come 
from a specific region of the main belt (see Table~\ref{tab:probSR}), i.e. the $\nu_6$ secular 
resonance, the 3:1, 5:2 and 2:1 Jupiter mean-motion resonances, the Hungaria region, the Phocaea region, and the
Jupiter Family Comets (JFC). 
\begin{table}[b]
    \centering
    \caption{The probability to come from each source region according to the model of
       \citet{granvik-etal_2018} for Kamo'oalewa. }
    \label{tab:probSR}
    \begin{tabular}{cc}
    \hline
    \hline
        Source region &  Probability   \\
    \hline
        $\nu_6$  &  0.73856  $\pm$  0.04940  \\
        3:1      &  0.05644  $\pm$  0.00016  \\
        5:2      &  0.00000  $\pm$  0.00000  \\
        Hungaria &  0.20491  $\pm$  0.04643  \\
        Phocaea  &  0.00001  $\pm$  0.00932  \\
        2:1      &  0.00009  $\pm$  0.00053  \\
        JFC      &  0.00000  $\pm$  0.00000 \\
    \hline
    \end{tabular}
\end{table}
Combining them with the NEOs albedo distribution by \citet{morbidelli-etal_2020}, 
an albedo distribution is first produced (see Fig.~\ref{fig:pdf_input}, left panel). 
The albedo $p_V$ is then converted into diameter $D$ with 
\citep[see, e.g.,][Appendix~A]{bowell-etal_1989, 2007Icar..190..250P} 
\begin{equation}
   D = \frac{1329 \text{ km}}{\sqrt{p_V}}10^{-H/5},
   \label{eq:mag2dia}
\end{equation}
using the absolute magnitude $H=24.3$ mag reported in NEODyS, assuming a typical uncertainty of 0.3 mag \citep{mainzer-etal_2011}.
The albedo distribution is also converted into density by defining three albedo categories, that we associated to the three asteroid complexes, namely C-, S-, X-complex. Specifically, we associated $p_V \leq 0.1$ to the C-complex, 
$0.1 < p_V \leq 0.3$ to the S-complex, and $p_V > 0.3$ to the X-complex.
The density of each asteroid complex is modeled with a lognormal distribution, using
the mean values and the standard deviations reported in Table~\ref{tab:astDensities}. 
\begin{table}[h]
    \centering
    \caption{The average density and the uncertainty of the three asteroid complexes.}
    \label{tab:astDensities}
    \begin{tabular}{ccc}
    \hline
    \hline
         Complex & Density (kg m$^{-3}$) & Uncertainty (kg m$^{-3}$)  \\
    \hline
            C    & 1200  & 300 \\
            S    & 2720  & 540 \\
            X    & 2350  & 520 \\
    \hline
    \end{tabular}
    \tablecomments{For the S-complex we used the density of S-type asteroids given in \citet{carry_2012}, while for C- and X-complexes we used the same values as \citet{fenucci-etal_2021}.}
\end{table}

The obtained probability density functions for Kamo'oalewa's diameter and density (Fig.~\ref{fig:pdf_input}, central and right panels) suggest that the most likely value of the diameter is around 32 meters, while the most likely value for the density is around 2450 kg m$^{-3}$.

\textbf{Thermal conductivity.} The thermal conductivity $K$ depends on the
composition, on the porosity, and on the surface temperature of the asteroid, and it can
vary of several orders of magnitude \citep{delbo-etal_2015}, depending on the nature of the
surface soil. In general, a low thermal conductivity in the range 0.0001 - 0.1 W
m$^{-1}$ K$^{-1}$ is associated to the presence of an insulating thermal layer on the
surface, composed by fine regolith grains, while moderate values in the range 
1-10 W m$^{-1}$ K$^{-1}$ are typical of a bare rock
composition.
This parameter is typically unknown, and additional data such as infrared observations or Yarkovsky 
effect detection are needed to attempt an estimation \citep{delbo-etal_2015}. 

To account for this uncertainty in K, we did not model its probability distribution, but rather used five reasonable values associated to different surface compositions. Namely, we used K $\in$  \{$0.001$, $0.01$, $0.1$, $1$, and $5$\} W~m$^{-1}$~K$^{-1}$, with the first three values accounting for the possibility of the presence of
regolith on the surface, and the last two representing a bare rock composition,
possibly with a different range of porosity. Note that the recent study by \citet{sanchez-scheeres_2020} has shown that small fast rotators could retain some regolith on their surfaces, and therefore the low thermal conductivity assumption for a small and super-fast rotator such as Kamo'oalewa is appropriate \citep[see also][]{fenucci-etal_2021}.
We did not take into account larger values indicative of an iron composition, since the 
spectrum obtained by \citet{reddy-etal_2017} suggests a classification as a S- or L-type, hence a stony composition.

\textbf{Heat capacity.} The heat capacity $C$ depends on the composition 
and on the surface temperature of the asteroid, and it might vary by a moderate factor.
Typical values assumed for regolith-covered or stony main belt asteroids are in the range
600-700 J kg$^{-1}$ K$^{-1}$ \citep{farinella-etal_1998}. 
For the purpose of this work, we assumed a slightly larger value of 800 J kg$^{-1}$ K$^{-1}$, 
since heat capacity increases with temperature and the surface of an asteroid placed at 
one astronomical unit from the Sun can reach temperatures as high as $\sim$400 K \citep{delbo-etal_2015}. 
Lower values typical of iron rich objects are excluded by the S- or L-type classification by
\citet{reddy-etal_2017}. We do not expect the final distribution of
$da/dt$ to be significantly different for heat capacity in the range 600-900 J
kg$^{-1}$ K$^{-1}$, since the largest differences are given by the uncertainty of the
thermal conductivity $K$.

\textbf{Obliquity.} The obliquity $\gamma$ of Kamo'oalewa is not known, hence we assumed
the obliquity distribution of the NEOs population derived by
\citet{tardioli-etal_2017}, that is given by
\begin{equation}
   p(\cos\gamma) = a \cos^2\gamma + b \cos\gamma + c,
   \label{eq:pdf_cosgamma}
\end{equation}
where the parameters are $a = 1.12, \, b = -0.32, \, c = 0.13$. 

\textbf{Rotation period.} The rotation period has been measured by
\cite{reddy-etal_2017}, and it is reported to be 0.467 hr. The entry of Kamo'oalewa in the
Asteroid Lightcurve Database\footnote{\url{http://alcdef.org/}} \citep[LCDB, ][]{warner-etal_2009}
indicates a quality factor of 2 for the lightcurve, meaning that the measured value might
be affected by an error of about 30 percent. We used therefore this value for the uncertainty and we model this parameter 
using a Gaussian distribution of the errors.

\textbf{Emissivity and absorption coefficient.} Measurements of the emissivity of meteorites are reported
in \citet{ostrowsky-bryson_2019}, and they are all between 0.9 and 1. Since we do not
expect the Yarkovsky drift to be significantly affected by the uncertainties in $\varepsilon$,
we assumed a fixed value, equal to $0.984$, that is the mean value of the measurements performed on meteorites.
The absorption coefficient $\alpha$ measures the fraction of sunlight absorbed by the surface of an asteroid, and a first approximation is given by $\alpha = 1 - A$, where $A$ is the Bond albedo \citep{bottke-etal_2006}. Typical values of asteroid's Bond albedo are smaller than 0.15 \citep[see e.g.][]{2001A&A...371..350V}, hence $\alpha$ varies between 0.85 and 1. Given the large uncertainties in the other parameters involved in the modeling of the Yarkovsky effect, we do not expect the absorption coefficient to play a significant role in the resulting semi-major axis drift distribution, hence we fix it to 1.

\subsubsection{Yarkovsky drift estimate for Kamo'oalewa}
\label{ss:yarkoEstimate}
\begin{figure}[b]
   \centering
   \includegraphics[width=0.47\textwidth]{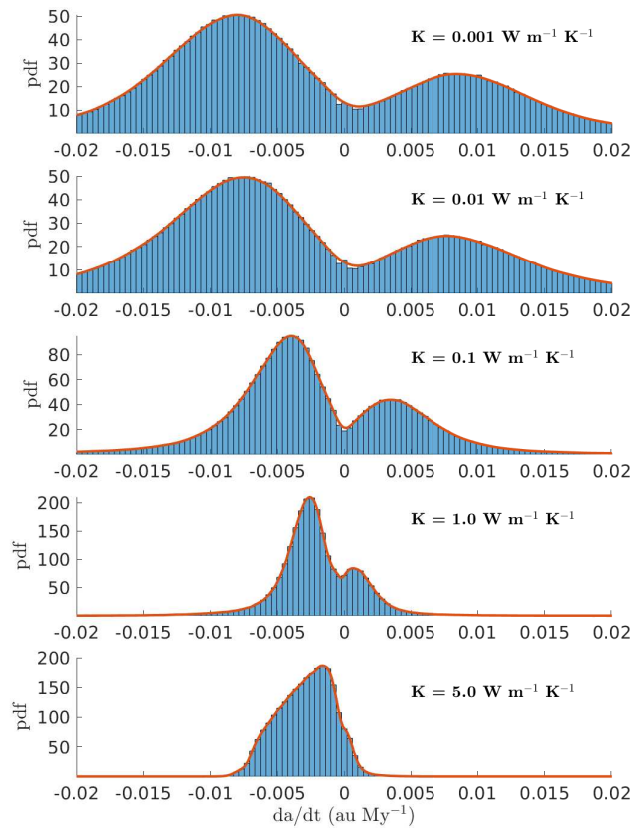}
   \caption{Modeled Yarkovsky effect induced semi-major axis drift rates $da/dt$ for asteroid Kamo'oalewa, given for different values of the thermal conductivity $K$, as indicated in the panels.
    }
   \label{fig:dadt2016HO3}
\end{figure}

Using the model described in Sec.~\ref{ss:parameterModeling}, we estimated the semi-major axis drift rate $da/dt$ of Kamo'oalewa by evaluating it from Eq.~\eqref{eq:yarkoForce} 
for a million different 
random combinations of the input parameters. A continuous probability density function
for the different values of thermal conductivity $K$ was then obtained using the
kernel density estimation, and the results are shown in
Fig.~\ref{fig:dadt2016HO3}.

For small and moderate thermal conductivity there are two most likely values of the semi-major axis drift rate, one negative and one positive, with the positive peak having a slightly
smaller probability than the negative one. This was however expected, due to
the assumed distribution of Eq.~\eqref{eq:pdf_cosgamma} for the obliquity $\gamma$, and the fact that the diurnal 
component, proportional to $\cos\gamma$, is the dominant one.
The peaks move towards smaller absolute values as $K$ increases,
and they finally
gather in a distribution with just one negative peak for $K=5$ W m$^{-1}$ K$^{-1}$. In this case the magnitude of the
seasonal component, that always produces a negative semi-major axis drift, is comparable to the magnitude of the diurnal one,
which results in a distribution with an excess of negative drifts. 
The fastest semi-major axis drifts are achieved for small conductivity, i.e. $K=0.001$ and $0.01$ 
W m$^{-1}$ K$^{-1}$, being the most likely values at around $\pm 0.008$ au My$^{-1}$,
while for $K=0.1$ W m$^{-1}$ K$^{-1}$ they are around $\pm 0.003$ au My$^{-1}$.
For the moderate conductivity $K=1$ and $5$ W m$^{-1}$ K$^{-1}$ the negative peak is
around $-0.002$ au My$^{-1}$, and positive values occur significantly less frequently.

\section{Results}
We produced 500 different orbital clones using the method of Sec.~\ref{ss:orbitalClones}, 
and integrated them using six different input distributions of
the Yarkovsky drift. The first run was performed in a purely gravitational model, while in the
remaining five we included the Yarkovsky effect assigning the drifts according to the
distributions obtained in Sec.~\ref{ss:yarkoEstimate}, and reported in Fig.~\ref{fig:dadt2016HO3}.
The orbits were propagated forward in time for 10 My
using a time step of 4 hours, while the orbital elements were recorded every 10 yr. 

\subsection{Long-term evolution}
\begin{figure}
    \centering
    \includegraphics[width=0.47\textwidth]{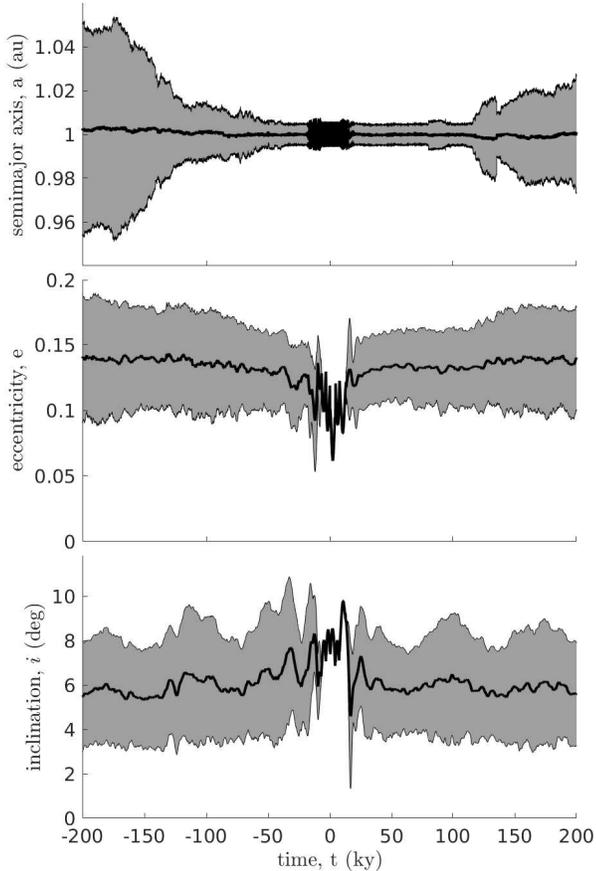}
    \caption{Evolution of orbital clones of Kamo'oalewa in the time interval (-200 ky, 200 ky), based on a purely gravitational model. The central thick black line represents the average of the orbital element, computed using the entire sample of 500 clones, while the gray filled area represents the corresponding standard deviation.}
    \label{fig:shortTermEvol}
\end{figure}

We considered first the evolution of the semi-major axis, eccentricity,
and inclination in the time interval (-200 ky, 200 ky), which is somewhat shorter than the My timescale.
We computed time-evolution of the average and the standard deviation of the elements. The result for the purely gravitational case is shown in Fig.~\ref{fig:shortTermEvol}.
In the vicinity of the starting epoch, the deviations are small, such that they can not
be appreciated by the eye in this scale; this means that all the clones have similar dynamics.
However, after a relatively short time of about 10 ky, the standard deviations grow 
significantly due to the chaotic dynamical nature of the NEO region, and the average value and the standard 
deviation are not anymore good representatives of the true dynamics. 
The introduction of the Yarkovsky effect produces small changes in the average values and the standard deviations computed in this time interval. The maximum difference in the average values is of the order of $\sim 5 \cdot 10^{-3}$ au in semi-major axis, $\sim 0.02$ in eccentricity, and $\sim 1^\circ$ in inclination, regardless of the input distribution used for the magnitude of the Yarkovsky drift, while the standard deviations remain comparable.    

\begin{figure}
    \centering
    \includegraphics[width=0.47\textwidth]{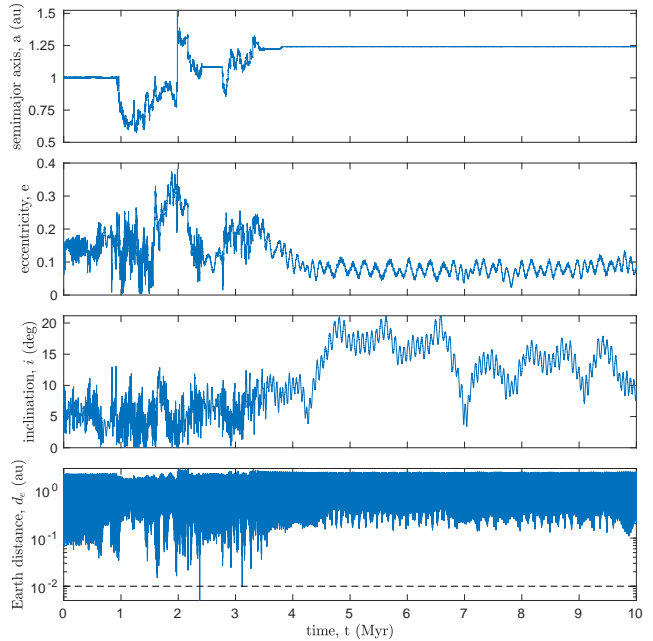}
    \caption{Long-term evolution of the nominal orbit forward in time for 10 My, computed in a purely gravitational model. From the top to the bottom, the panels show the semi-major axis, eccentricity, inclination, and the distance from the Earth. The dashed line in the bottom panel indicates the radius of the Hill sphere of the Earth.}
    \label{fig:longNominal}
\end{figure}

The long-term evolution of semi-major axis, eccentricity, and inclination for the nominal orbit in a purely gravitational model is shown in 
Fig.~\ref{fig:longNominal}, together with the evolution of the distance from the Earth. The asteroid is removed from
the 1:1 mean motion resonance with the Earth after about 1 My, after which it suffers several close encounters 
with the Earth during 2.5 My, while the eccentricity does not exceed 0.4 and the inclination is smaller than 10$^\circ$. 
At about 3.5 My the object ends up in a stable almost circular orbit at about 1.25 au, corresponding to the 7:5 mean motion 
resonance with the Earth, and it remains there until the end of the 10 My integration. However, the analysis of the nominal orbit 
on its own is not enough to assess the long-term stability, that needs to be studied statistically using orbital clones.

To study the effect of the introduction of the Yarkovsky drift in the long-term evolution of the 
clones, we computed
the time spent in the Earth co-orbital region $T_{1:1}$ 
and the orbital distribution $R(a,e,i)$ of the clones. 
For the definition of the Earth co-orbital region, we used the inequalities 
\begin{equation}
    0.994 \text{ au} \leq a \leq 1.006 \text{ au}
    \label{eq:earthCOORBregion}
\end{equation}
given by \citet{delafuente-delafuente_2016}.
\begin{figure*}[!ht]
    \centering
    \includegraphics[width=\textwidth]{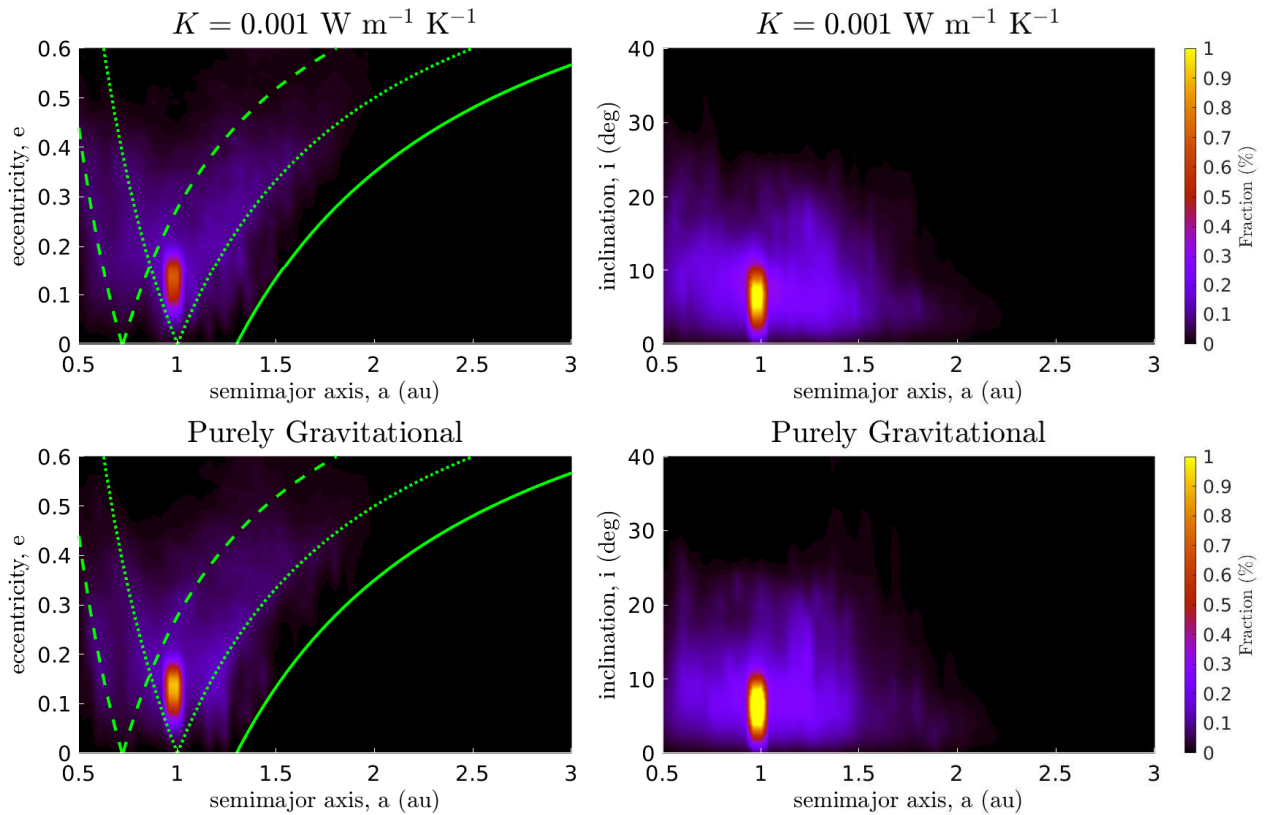}
    \caption{Orbital distributions of the clones, in the planes $(a,e)$ (left column) and $(a,i)$ (right column),
    obtained for $K=0.001$ W m$^{-1}$ K$^{-1}$ (top row), and in a purely gravitational model (bottom row). In the $(a,e)$ plots, the thick green curve is the boundary of the NEO region $q = 1.3$ au, the dotted green curves are $q=q_{\text{Earth}}$ and $Q=Q_{\text{Earth}}$, and the dashed green curves are $q=q_{\text{Venus}}$ and $Q=Q_{\text{Venus}}$, where $q,Q$ are the perihelion and aphelion distance, respectively.}
    \label{fig:ODgrav}
\end{figure*}
To compute the orbital distribution $R(a,e,i)$, we divided the space of the orbital elements $(a,e,i)$ 
in cells of volume $\Delta a \times \Delta e \times \Delta i$, where 
$\Delta a = 0.04 \text{ au}, \ \Delta e = 0.01, \ \Delta i = 1^\circ$, 
and we recorded the total time spent in each cell. 
Then, we normalized the distribution so that
\begin{equation}
   \iiint R(a,e,i) \text{d}a \, \text{d}e  \, \text{d}i = 1.
   \label{eq:normalizedOD}
\end{equation}
In practice, the orbital distribution gives information on how much time the clones 
spent in a given region of the orbital elements space.

The orbital distributions of the clones obtained for $K=0.001$ W m$^{-1}$ K$^{-1}$, and in a purely gravitational model are shown in Fig.~\ref{fig:ODgrav}.
The most dense area is a strip along the Earth co-orbital zone at $a\sim1$ au at inclinations smaller than 10$^\circ$ and eccentricity smaller than 0.2, meaning that clones spent a significant fraction of their evolution here. It can also be noted that when the Yarkovsky effect is introduced, the density on this strip is slightly reduced.
Outside this strip, clones are spread throughout the NEO region (and occasionally also beyond) by effect of close encounters with the Earth and Venus. However, most of the clones keep an eccentricity smaller than roughly 0.5 and an inclination smaller than 30$^\circ$. The introduction of the Yarkovsky effect in the model seems to produce small changes in the orbital distributions, that are difficult to quantify due to small number statistics outside the strip at $a \sim 1$ au.

\begin{figure}
    \centering
    \includegraphics[width=0.48\textwidth]{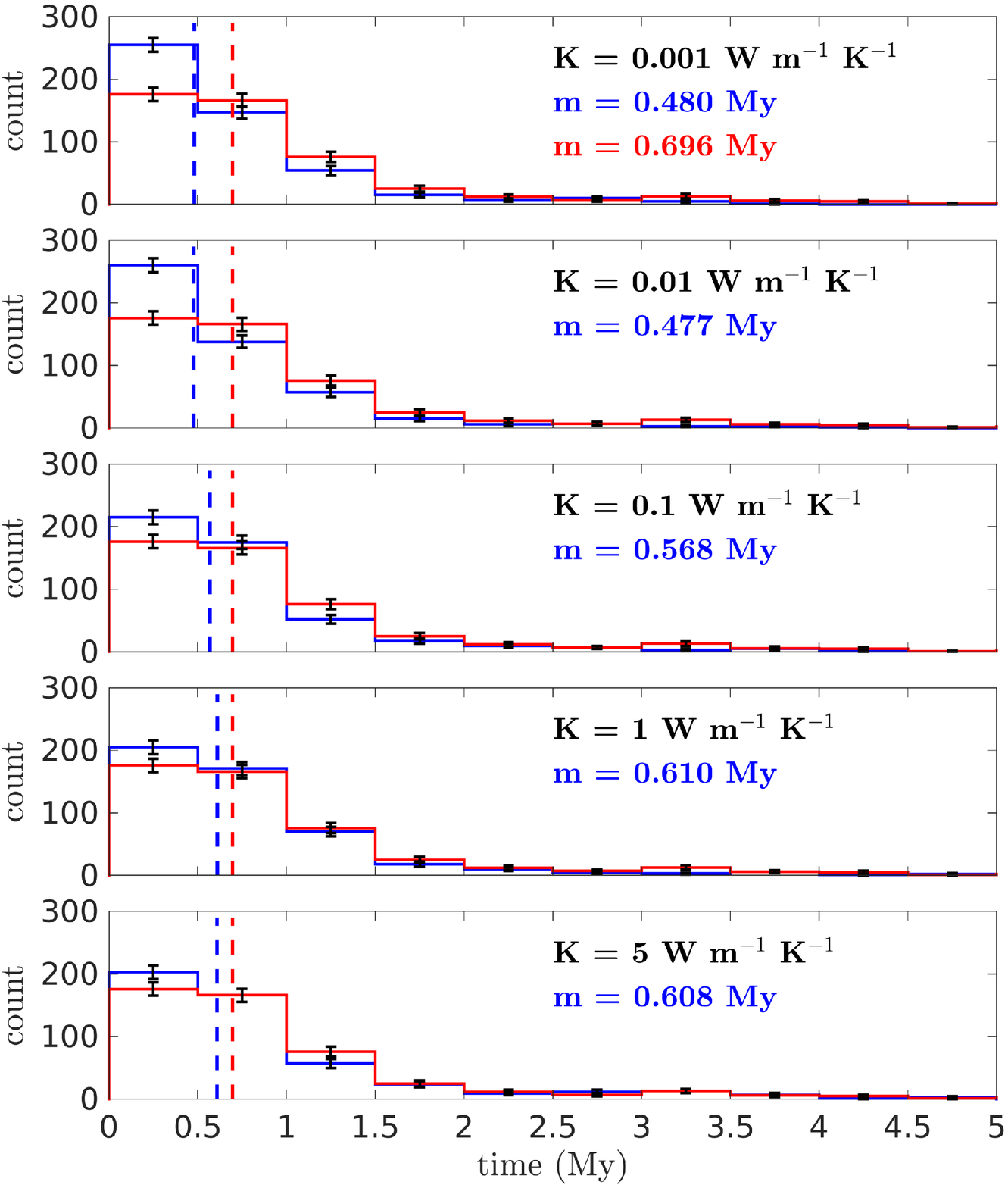}
    \caption{Distributions of the time spent in the Earth co-orbital region $T_{1:1}$. The blue color refers to results obtained with the Yarkovsky effect included in the model, and the corresponding value of conductivity used in the simulations is reported in each panel. The red color refers to results obtained in a purely gravitational model, that are shown for comparison. Vertical dashed lines indicate the median of the distributions, and the corresponding numerical values are written in each panel. Error bars in the histograms are relative to each entire bin, and they are computed by using Bernoulli statistics.
    }
    \label{fig:T11distrib}
\end{figure}

The distributions of the time spent in the Earth co-orbital zone $T_{1:1}$, computed with the 
introduction of the Yarkovsky effect in the model, as well as in a purely gravitational model, are shown in Fig.~\ref{fig:T11distrib}, using 0.5 My width for the bins. 
A vast majority of the clones escaped the Earth co-orbital region in less than 2.5 My, 
regardless whether or not the Yarkovsky effect is included in the model, and objects that survive longer are quite rare. 
Still, in the cases $K = 0.001$ and $0.01$ W m$^{-1}$ K$^{-1}$ the introduction of the semi-major axis drift in the dynamics caused clones to be removed faster from the Earth co-orbital region than in a purely gravitational model. This can be best appreciated from the first bin of the histograms shown in Fig.~\ref{fig:T11distrib}. 
On the other hand, the histograms obtained with $K = 0.1, \, 1$ and $5$ W m$^{-1}$ K$^{-1}$ almost agree within error bars with the histogram computed using a purely gravitational model. This indicates that a less efficient Yarkovsky drift, caused in our case by a high thermal conductivity, produces smaller changes in the distribution of $T_{1:1}$. Figure~\ref{fig:T11distrib} also shows that the median of the distributions move towards larger values as the thermal conductivity increases, passing from about 500 ky in the cases $K = 0.001, \, 0.01$ W m$^{-1}$ K$^{-1}$, to about 550 ky in the case $K = 0.1$ W m$^{-1}$ K$^{-1}$, and finally at about 600 ky in the case $K = 1$ and $5$ W m$^{-1}$ K$^{-1}$. As a comparison, the median for the purely gravitational model was 690 ky.

The distributions of the time spent in the Earth's co-orbital region could be fitted well with log-normal distributions, i.e. with random variables of the form $X = e^{\mu + \sigma Z}$, where $Z$ is the standard normal variable. Note that moving $\mu$ to lower values while keeping $\sigma$ constant shifts the peak of the log-normal distribution to lower values and increases the peak probability.

\begin{figure}
    \centering
    \includegraphics[width=0.47\textwidth]{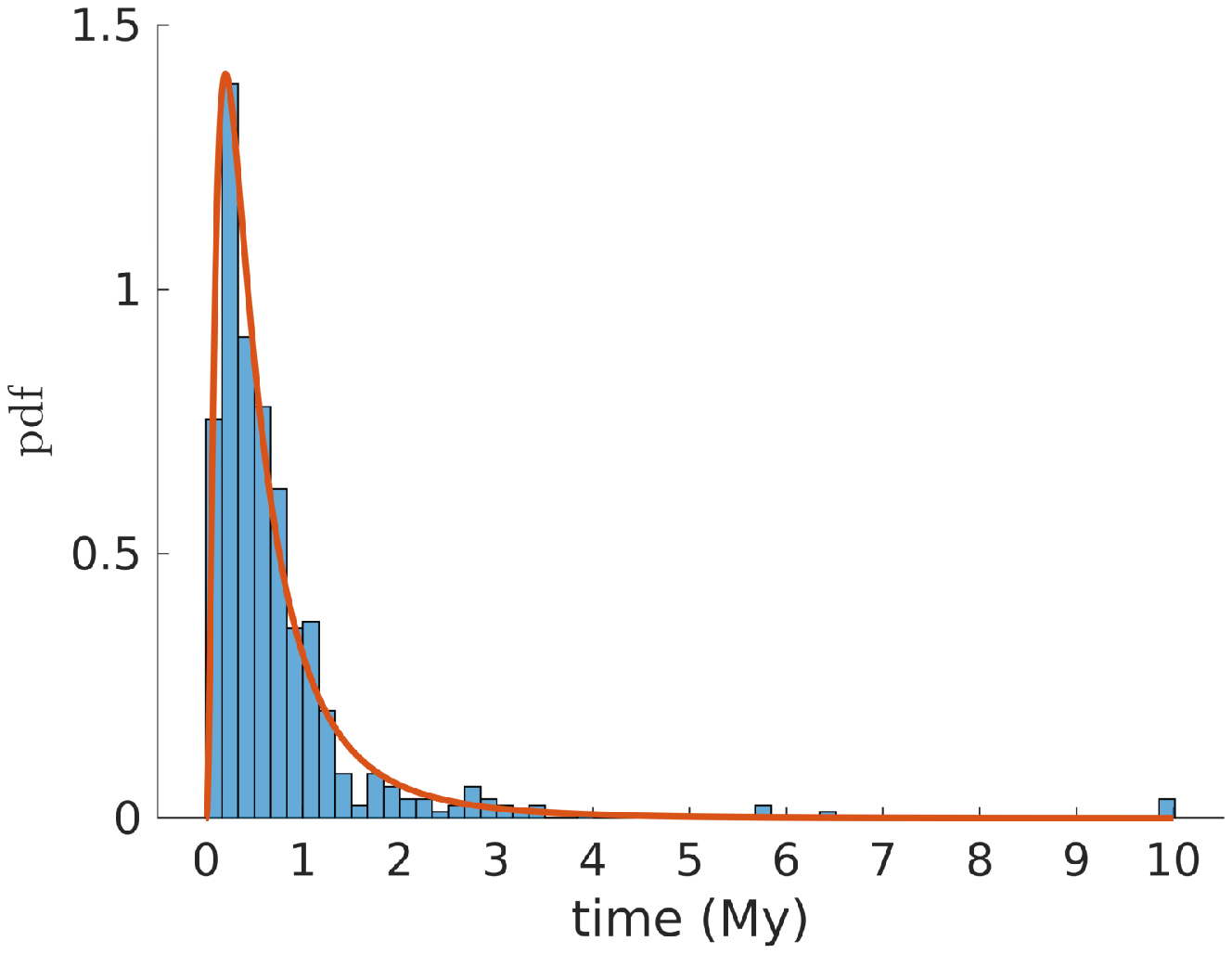}
    \caption{Fit of the distribution of $T_{1:1}$ with a log-normal distribution for the case $K=0.001$ W m$^{-1}$ K$^{-1}$.}
    \label{fig:lognormalFit}
\end{figure}

Taking advantage of this, we further analyze how the magnitude of the Yarkovsky effect correlates with the time $T_{1:1}$. Figure~\ref{fig:lognormalFit} shows the fit obtained for $K=0.001$ W m$^{-1}$ K$^{-1}$, together with the histogram of $T_{1:1}$ with 60 uniform bins. By using an increased number of bins, we can appreciate that a log-normal distribution is a good model for the $T_{1:1}$ distribution. Similar figures have been obtained for all the other simulations we performed, including the case of the purely gravitational model.

\begin{figure*}
    \centering
   \includegraphics[width=0.95\textwidth]{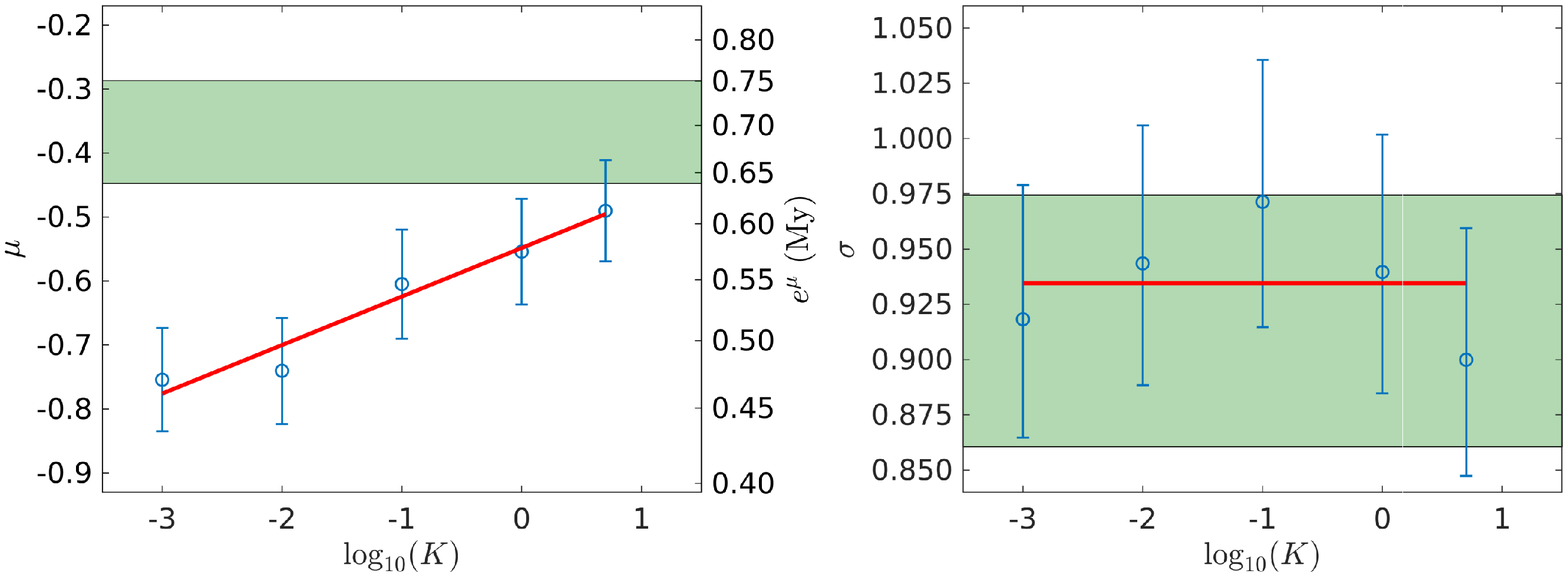}
    \caption{Estimated values of the parameters $\mu$ (left panel) and $\sigma$ (right panel), together with their linear trend in red. The green area represents the estimated range of $\mu$ and $\sigma$ for the purely gravitational case.}
    \label{fig:mu_sigma_fits}
\end{figure*}

Figure~\ref{fig:mu_sigma_fits} shows the parameters $\mu$ and $\sigma$ of the log-normal distribution for the five different values of thermal conductivity, together with their estimated uncertainties. The parameter $\mu$ follows a clear linear trend in thermal conductivity. As a comparison, the estimated value of $\mu$ in the case of the purely gravitational model is $\mu = -0.367 \pm 0.080$, and this range is statistically compatible with the simulations that included the Yarkovsky effect only when $K = 5$ W m$^{-1}$ K$^{-1}$. On the other hand, though $\sigma$ also shows some variations, the obtained results are still statistically indistinguishable. They are all compatible with the value of $\sigma = 0.914 \pm 0.053$, estimated from the purely gravitational model. 
Note that $e^\mu$ corresponds to the median value of the log-normal distribution, so that the scale used on the right axis of the left panel of Fig.~\ref{fig:mu_sigma_fits} can be used for a comparison with the values reported in Fig.~\ref{fig:T11distrib}.
Combining the results obtained for $\mu$ and $\sigma$, and considering how the log-normal distribution depends on these parameters, provides additional evidence that clones are removed faster from the Earth co-orbital region for low thermal conductivity.

In the low thermal conductivity case, the Yarkovsky effect causes statistically significant changes in the time Kamo'oalewa spends in the Earth co-orbital region.
On the contrary, for moderate and high values of the thermal conductivity, the produced changes are at the level of the shot noise. Nevertheless, the possibility for this asteroid to remain an Earth co-orbital object for more than 1 My, as proposed by \citet{delafuente-delafuente_2016}, seems unlikely, since about 80
percent of the clones exited from this region in less than 1 My. Similarly, the time needed to escape from the 1:1 mean motion resonance is shorter as the assumed thermal conductivity $K$ decreases. Therefore, if the magnitude of the Yarkovsky effect is large, it may significantly affect the orbital motion of Kamo'oalewa. Such Yarkovsky-induced drift rates are possible only if the thermal conductivity is low. In this respect, we recall that a low surface thermal conductivity scenario is feasible for the super-fast rotators, as shown in \citet{fenucci-etal_2021}. 

\subsection{Collisions with planets}
The \texttt{mercury} integrator is also capable to record collisions with the Sun and planets. 
Clones that did not survive the whole 10 My integration were either expelled from the Solar System,
or collided with the Sun or a planet. 
We considered an object to be expelled from the Solar System if its semi-major axis exceeded 100 au at some point 
of the evolution. The statistics of end-up states for all the clones is given in Table~\ref{tab:coll}.

The majority of the clones ended their orbit evolution colliding with the Sun or, more importantly, with Venus or the Earth. About 6-9 percent of the objects hit the Sun, while the collision probability with Venus and the Earth is roughly constant at about 12-15 percent. This means that there is no statistically significant trend in the collision probabilities with the thermal conductivity. Only 1-4 clones for each run of simulations ended up in a collision with Mercury, while collisions with Mars happen even more rarely.

\begin{table*}[!htbp]
    \centering
    \caption{The number of objects that ended their evolution due to a collision with the Sun, Venus or the Earth, for the different distributions of semi-major axis drift obtained using the fixed value of $K$ given in the first column. The last but one column reports the number of objects that have been expelled from the Solar System or that collided with any other planet, while the last column reports the number of objects that survived the whole 10 My integration timespan.}
    \label{tab:coll}
    \begin{tabular}{cccccc}
    \hline
    \hline
    $K$ (W m$^{-1}$ K$^{-1}$) & Coll. with Sun & Coll. with Venus & Coll. with the Earth 
    & Other removed objects & Survived objects \\
    \hline
    0.001               & 45 & 55 & 73 & 10 & 317 \\
    0.01                & 27 & 72 & 70 & 12 & 319\\
    0.1                 & 42 & 72 & 69 & 8  & 309\\
    1                   & 44 & 60 & 52 & 9  & 325\\
    5                   & 43 & 79 & 68 & 16 & 305\\
    \textbackslash      & 34 & 65 & 65 & 10 & 326\\
    \hline
    \end{tabular}
\end{table*}

Only 1-2 percent of the clones have been ejected from the Solar System due to close encounters with Jupiter. The number of objects hitting the planet is even smaller, only 2 over 3000 total clones. Collisions with the outer planets, as well as escapes from the Solar System, are infrequent. This is because close encounters with inner planets are, in general, not strong enough to put an object in a hyperbolic trajectory. Additionally, there is no dynamical effect that moves the clones close to Jupiter, Saturn, or Neptune. Also, all the surviving clones remained inside the NEO region for the whole 10 My integration timespan. Therefore, these results suggest that the Yarkovsky effect does not change the statistics of the ending states of the clones.

\section{Discussion} \label{sec:discussion}

\subsection{Limitations of the Yarkovsky model}
The Yarkovsky effect used in our simulations is implemented as a constant secular 
acceleration along the orbital velocity (see Sec.~\ref{ss:numericalModel}). 
This simplistic approach however has some limitations. Two most important issues
are orbit and spin axis evolution. 

\subsubsection{Orbit evolution}

Regarding the orbit evolution, the Yarkovsky effect depends on the heliocentric distance, i.e. on the semi-major axis and eccentricity of the orbit
\citep[see, e.g.,][]{bottke-etal_2006,vokrouhlicky-etal_2015}, becoming more efficient when the object moves closer to the Sun. Therefore, as the orbital elements evolve, 
the magnitude of the drift should evolve as well.

However, we found that the Yarkovsky effect mainly affects the residence time in the Earth co-orbital zone, where $a$ is bounded within the interval given by Eq.~\eqref{eq:earthCOORBregion}. This strongly limits variations in the semi-major axis, hence the orbit evolution should not produce any change in the results obtained assuming a constant Yarkovsky induced drift.
Another factor that may change the magnitude of the drift is the orbital eccentricity. The distributions of $da/dt$ 
of Sec.~\ref{ss:yarkovskyDrift} are computed using a model of the Yarkovsky effect that assumes a circular
heliocentric orbit for the asteroid. 
However, NEOs can reach very high eccentricity that makes the Yarkovsky effect more efficient. In particular, \citet{spitale-greenberg_2001} found that for 10 to 100 meters sized asteroids the diurnal Yarkovsky 
effect can increase by more than about an order of magnitude when the eccentricity passes from $\sim$0.1 to $\sim$0.9.
In this case, our constant drift assumption may be inappropriate.
Still, the orbital distributions of Fig.~\ref{fig:ODgrav} suggests that only a small
fraction of clones ended up in highly eccentric orbits. This can also be noticed in the number of collisions
with the Sun and escapes from the Solar System (see Table~\ref{tab:coll}), 
which represented only 10 percent of the total number of clones. 
Moreover, Fig.~\ref{fig:ODgrav} shows that clones presumably exit the Earth 
co-orbital region with an eccentricity smaller than 0.3. Therefore, the increase of the magnitude of the semi-major axis drift due to high eccentricity should not significantly
change the results obtained in our analysis, especially regarding the residence time in the Earth co-orbital region.

\subsubsection{Spin axis orientation}

Another issue of our Yarkovsky model presented in Sec.~\ref{ss:yarkovskyDrift} is that it assumes a constant orientation of the spin axis. Different factors could change the rotational state of an asteroid, such as: the Yarkovsky-O'Keefe-Radzievskii-Paddack (YORP) effect \citep{2000Icar..148....2R,bottke-etal_2006,vokrouhlicky-etal_2015}, meteoroid impacts \citep{2015Icar..252...22W}, and planetary close encounters \citep{2005Icar..178..281S,2018A&A...617A..74S,2020EPJST.229.1391B}. 

While the spin evolution of an asteroid not subjected to collisions is expected to be dominated by the YORP effect, collisions could also change the spin parameters. Meteoroids' impacts could be significant enough to randomly re-orient the spin axis, and \citet{2015Icar..252...22W} suggested that their effect on the rotational state could be as important as the YORP effect, especially for asteroids of a few tens of meters in radius. 

However, typical timescales for spin axis re-orientation due to collisions in the main-belt are given in \citet{farinella-etal_1998}, and they depend on the asteroid size and rotation period. The timescale found for a Kamo'oalewa-like object is of the order of 10 My, comparable to our
total integration timespan and well above the median time of 0.5 My needed to exit from the 1:1 resonance with the Earth. Furthermore, the main contribution of the Yarkovsky effect on the dynamics is a faster removal from the Earth co-orbital zone, which occurs mostly in the first 1 My of the dynamical evolution, and even more probably in the first half My.
In the NEO region the timescale for impact-induced spin axis re-orientation should be even longer, and this mechanism should not play an important role for our model.  
Similarly, although in principle close planetary encounters could also change the spin rate and the orientation of the spin axis, in our simulations the orbital clones are placed onto planetary-crossing orbits only once they escape from the 1:1 mean motion resonance due to the growing eccentricity. Consequently, close encounters may affect spin states only in late phases, also making these effects less important for our conclusions.

Evaluating the importance of the YORP effect is a much more complicated issue. Therefore, we gave it special attention and performed additional simulations to understand better its role.
The YORP effect is a thermal torque that pushes the obliquity $\gamma$ towards one of the asymptotic states (0, 90, or 180 degrees), while the asteroid rotation is sped up or slowed down. However, its effect sensitively depends on the object's shape and thermal properties, which are not known in the case of the Kamo'oalewa. 
For this reason, we adopted a strategy that circumvents the above-described problem but still allows us to reasonably assess the importance of the YORP effect on our results. Our methodology is based on a simplified model of the YORP effect and a statistical approach, rather than deterministic. We also focused on the cases with thermal conductivity  $K\leq0.01$ W m$^{-1}$ K$^{-1}$, for which we found the Yarkovsky effect to be important in the dynamics of Kamo'oalewa.

Denoting with $\omega$ the rotation rate, \citet{2004Icar..172..526C} computed the vector field $(\text{d}\gamma/\text{d}t, \text{d}\omega/\text{d}t)$ for a sample of 200 shapes in the case $K=0.001$ W m$^{-1}$ K$^{-1}$ (see Fig.~7 therein). To account for the unknown shape of Kamo'oalewa, for each clone, we randomly select the vector field among those provided by \citet{2004Icar..172..526C} and keep it for the whole integration timespan.  Furthermore, we allow the evolution of $\text{d}\gamma/\text{d}t$ in such a way that 0 and 180 deg are the only possible asymptotic states for the obliquity.
The current observed fast rotation of Kamo'oalewa may indicate that the YORP effect is speeding up this object. Hence we selected $\text{d}\omega/\text{d}t$ so that the asymptotic obliquity can be reached only by decreasing the rotation period. 

To the purpose of these simulations, we implemented the spin axis evolution in the \texttt{mercury} integrator similarly to \citet{2011MNRAS.414.2716B}. This implementation allows to update the semi-major axis drift $\text{d}a/\text{d}t$ according to the changes in obliquity $\gamma$ and rotation rate $\omega$.
The initial values for density, diameter, and obliquity are generated as described in Sec.~\ref{ss:parameterModeling}. The initial rotation period is assumed to be $P=6$ h, and the thermal conductivity is fixed at $K=0.001$ W m$^{-1}$ K$^{-1}$.

The described model is used to simulate the evolution of 500 Kamo'oalewa's orbital clones, that are propagated over 10 Myr. About 65 percent of the clones reached 180 deg in obliquity, while the remaining fraction reached 0 deg. The median time needed to arrive at the asymptotic state was only about 2.3 kyr, and 80 percent of the clones reached it in less than 10 kyr. The median time needed to achieve the current rotation rate was about 230 ky, and about 80 percent of the clones reached it in less than 1 My.

The lognormal fit of the values of $T_{1:1}$ gave the parameters $\mu = -0.754 \pm 0.08, \, \sigma = 0.918 \pm 0.06$, which are compatible at 1$\sigma$ level with the results obtained with a constant Yarkovsky effect in the cases $K = 0.001$ W m$^{-1}$ K$^{-1}$ (see Fig.~\ref{fig:mu_sigma_fits}). This suggests that the YORP effect should not change significantly the results obtained assuming a constant semi-major axis drift.

However, caution is needed because our simplified YORP model does not account for some evolution pathways. \citet{2004Icar..172..526C} found that about 20 percent of objects with $K=0.001$ W m$^{-1}$ K$^{-1}$ achieve 90 deg as an asymptotic state by slowing down the rotation rate. At this obliquity, the diurnal component of the Yarkovsky effect is switched off, and only the seasonal component is operating. Similarly, when the rotation period is 
very long the Yarkovsky effect is again minimized.
In such cases, the magnitude of the semi-major axis drift would probably be too small to significantly affect the dynamics of Kamo'oalewa. Still, the above scenario seems less likely in the case $K = 0.01$ W m$^{-1}$ K$^{-1}$, where \citet{2004Icar..172..526C} found that only 5 percent of asteroids reach 90 deg as an asymptotic state.

Yet, the spin axis dynamics could have more features. \citet{2021AJ....162....8G} found that there is about the same probability to reach 0/180 deg and 90 deg in obliquity for low thermal inertia. Moreover, their model presents additional attractive equilibrium points \citep[as already seen in e.g.][]{2008CeMDA.101...69S}. The authors suggested that after several YORP cycles, the spin state could reach such an equilibrium point and remain locked. The simulations performed for Kamo'oalewa show that the asymptotic states are achieved in a few thousand years, indicating that such an equilibrium state could be reached within just the first $\sim$50 ky of the evolution. In this case, our assumption of a constant semi-major axis drift $\text{d}a/\text{d}t$ would be fully justified, though the magnitude of the drift would depend on the equilibrium point reached.

Furthermore,
\citet{2015ApJ...803...25C} investigated the effects of shape changes induced by the spin evolution, and found self-limitating phenomena. For instance, the spin axis may undergo a random-walk evolution (stochastic YORP). There these events could remove an object from an equilibrium, making the spin axis evolving again.
Another possibility is that the object toggles between a small number of configurations, causing $\gamma$ and $\omega$ to evolve within a narrow range (self-governing YORP). 

Therefore, in many possible scenarios, Kamo'oalewa's spin axis evolution due to the YORP effect should not significantly reduce the magnitude of the Yarkovsky effect. In these cases, our conclusions about the importance of the Yarkovsky effect in the dynamics of Kamo'oalewa are still valid. However, we must recognize that different scenarios also exist, and that the influence of the YORP effect on our results is uncertain.
These uncertainties will be possible to eliminate only once the shape and the physical properties of Kamo'oalewa are reliably determined.

\subsection{Connection with Barbarian asteroids}
As discussed in Sec.~\ref{sec:basic}, Kamo'oalewa is possibly an L-type asteroid. This taxonomic class is linked to the so-called Barbarian asteroids, which are characterized by unusual polarimetric behavior. 
\citet{2018Icar..304...31D} modeled the spectra of L-type objects and found that it can be successfully explained using mixtures of the spinel present in Calcium Aluminum-rich Inclusions (CAIs), MgO-rich olivine, and the mineral compounds found in the matrix of CV3 meteorites. The authors also concluded that the presence of CAIs is responsible for the abnormal polarimetric inversion angle of the Barbarians. In this respect, Barbarians could be among the oldest main-belt asteroids, containing the samples of the oldest solid materials found in the Solar system \citep{2008Sci...320..514S,2019MNRAS.485..570C}.

If confirmed to be an L-type object, the origin of Kamo'oalewa may be linked to existing Barbarians in the main asteroid belt. Recently, \citet{kuroda-etal_2021} identified the first Barbarian asteroid in the NEO population.
The main known reservoirs of these objects are the Watsonia \citep{2014MNRAS.439L..75C} and Brangane \citep{2019MNRAS.485..570C} asteroid families. While the origin of Kamo'oalewa in the high-inclination Watsonia family \citep{2011Icar..216...69N} seems unlikely, it might be arriving from the Brangane family via the 3:1 mean motion resonance with Jupiter (see Table~\ref{tab:probSR}). Also, despite there is no collisional family associated with the first known Barbarian asteroid (234)~Barbara \citep{2014Icar..239...46M}, Kamo'oalewa might still be its fragment. A more in-depth analysis is needed to possibly establish a strong dynamical link between Kamo'oalewa and Barbarian asteroids.

Barbarians are extremely rare and ancient and were present before Earth's formation. A possible link of Kamo'oalewa to this class of objects would make it probably the best possible target for the next sample return mission to an asteroid. Therefore, it is of the highest priority to extend the determination of the spectra beyond 1~$\mu$m, to confirm its L-type classification.

\section{Conclusions}
In this paper, we studied the long-term dynamics of asteroid (469219) Kamo'oalewa by integrating orbital clones forward in time for 10 My and taking into account the semi-major axis drift produced by the Yarkovsky effect. Our simulations showed that the Yarkovsky effect plays a significant role only in the case of low surface thermal conductivity, and it causes a faster removal from the Earth co-orbital zone. Many clones were removed in the first 0.5 My, and about 80 percent of them exited during the first 1 My of evolution. 

There is about 12-15 percent probability that Kamo'oalewa will end up colliding with the Earth. Moreover, we found that these collision probabilities do not depend on the Yarkovsky drift rates used in our simulations. Although the semi-major axis drift alters the residence time in the Earth co-orbital region, Kamo'oalewa is still the most stable Earth co-orbital object known so far.

\acknowledgments

We thank the anonymous referee for the comments and suggestions that helped us to improve the manuscript. The authors have been supported by the MSCA-ITN Stardust-R, Grant Agreement n. 813644 under the European Union H2020 research and innovation program. BN also acknowledges support by the Ministry of Education, Science and Technological Development of the Republic of Serbia, contract No. 451-03-9/2021-14/200104.

\bibliography{stardustBib}{}
\bibliographystyle{aasjournal}



\end{document}